\documentclass[a4paper]{scrartcl}

\usepackage[utf8]{inputenc}
\usepackage[english]{babel}
\usepackage[T1]{fontenc}

\usepackage[inline]{enumitem}

\usepackage{booktabs}

\usepackage{mathtools, amssymb}
\usepackage{ntheorem}
\usepackage{bm, dsfont}

\usepackage[style=numeric, url=true, maxnames=4, sortcites=true]{biblatex}
\usepackage{todonotes}
\usepackage{graphicx}
	\graphicspath{{experiments/figures/}}

\definecolor{aaublue}{rgb}{0.00,0.41,0.66}
\definecolor{nicered}{rgb}{0.647,0.129,0.149}

\usepackage{hyperref}
\hypersetup{%
	breaklinks=true,%
	colorlinks=true,%
	linkcolor=blue,%
	citecolor=nicered,%
	urlcolor=aaublue,%
	pdftitle={},%
	pdfauthor={RDJ, JM, MN, MGR},%
	pdfsubject={generalized sampling, point processes},%
	pdfkeywords={generalized sampling, point processes},%
	pdfpagemode=UseOutlines,%
	pdfstartview=FitH%
}

\usepackage[capitalise, poorman]{cleveref}
	\crefname{figure}{Figure}{Figures}
	\crefname{enumi}{}{}
	\crefformat{equation}{(#2#1#3)}
	\crefname{equation}{}{}

\newcommand\cH{\ensuremath{\mathcal{H}}}

\DeclarePairedDelimiterX\Set[1]{\{}{\}}{
	#1
}
\newcommand{\SetLetter}[1]{\ensuremath{\mathbb{#1}}}
\newcommand{\N}{\SetLetter{N}}

\newcommand{\R}{\SetLetter{R}}

\newcommand\ci{i}

\DeclarePairedDelimiter\abs{\lvert}{\rvert}

\DeclareMathOperator{\spn}{span}

\DeclarePairedDelimiterX\inner[2]{\langle}{\rangle}{#1, #2}

\DeclarePairedDelimiter\norm{\lVert}{\rVert}

\newcommand{\T}[1]{\top}

\newcommand{\bbbone}{\ensuremath{\mathds{1}}}

\DeclareMathOperator*{\argmin}{arg\,min}

\newcommand\pval{\ensuremath{\text{p-value}}}

\theoremheaderfont{\normalfont\bfseries}
\theoremstyle{break}

\newtheorem{theorem}{Theorem}
\theorembodyfont{\normalfont}
\newtheorem{definition}[theorem]{Definition}

\title{Investigations of the effects of random sampling patterns on the stability of generalized sampling} 

\author{Robert Dahl Jacobsen\thanks{Department of Mathematical Sciences, Aalborg University, Fredrik Bajers Vej 7G, DK-9220 Aalborg East} 
\and Jesper Møller\footnotemark[\value{footnote}]{} 
\and Morten Nielsen\footnotemark[\value{footnote}]{} 
\and Morten Grud Rasmussen\footnotemark[\value{footnote}]{}}

\begin{document}

\maketitle

\begin{abstract}
    Generalized sampling is a mathematical technique for obtaining approximations of signals with respect to different representations in a numerically stable manner.
    This can for example be relevant in processing MRI images, where hardware often enforces initial frequency measurements, but where a wavelet basis may be better suited for representing the image.

    Recently the theory of generalized sampling was extended to work with arbitrary patterns in $\R^d$.
    In this article we investigate how the choice of the probability distribution generating random sampling schemes in $\R^2$ affects the numerical stability of generalized sampling.
\end{abstract}

\section{Introduction}

Generalized sampling \cite{Adcock:Hansen:2012,Adcock:2017} is a technique for computing a representation of a function in one basis/frame of a Hilbert space from another.
The theory is abstract and does not restrict the type of bases or frames that can be considered, and this provides the freedom to adapt the setup to the structure of the functions at hand.
In a typical application, we have samples of the function given as inner products with respect to some fixed basis or frame imposed by the measuring process, but prior knowledge of the general function structure dictates a desire to change to another more efficient representation system for the function.
For example,  one may have access to Fourier samples of an image, but would like to change to a more efficient representation system for images such as wavelets.

An essential quantity ensuring numerical stability of the generalized sampling approach is the condition number of the change of basis (or frame) matrix (for short ``condition number'') between the two representation systems considered, see \cite{Adcock:2017}.
The condition number will, in general, depend on the sampling scheme and on the sampling and representation system.
If the sampling scheme is a subset of a regular grid, the numerical stability of the change of basis matrix is well understood in the Fourier/wavelet setup \cite{Adcock:Hansen:Kutyniok:Ma:2014,MR3116643,MR3175085}.
However, the data acquisition process often forces one to consider non-regular sampling schemes \cite{Ahn:1986} where numerical stability is less well understood.

In this paper we consider sampling schemes obtained randomly in the Fourier/wavelet setup and study how the condition number can be controlled by using binomial, Poisson and determinantal point processes (DPPs, see \cite{Lavancier:Moeller:Rubak:2014,Soshnikov} and the references therein).
We demonstrate that DPPs are particularly useful because nearby points in the process repel each other.
There are previous studies of the numerical stability of Fourier systems with random sampling schemes \cite{Bass:Groechenig:2005,Rauhut:2007}, but only using the binomial point process.

The condition number depend on the density and bandwidth of the sampling scheme (as defined in \cref{sec:NUGS}). 
For the Poisson process, we derive a theoretical result (\cref{t:5}) relating the density and the intensity of the process.
Estimating the probability that a random point pattern satisfies an appropriate density criterion is a natural extension of results in \cite{Adcock:2017} for the parameters of non-random sampling schemes.

\section{Generalized sampling}
\label{sec:NUGS}

In the following, we recap the framework of generalized sampling, introduced in a series of papers by Adcock, Hansen and collaborators, see \cite{Adcock:Hansen:2012,Adcock:2014,Adcock:2017}.
Let $\cH$ be a separable Hilbert space and let $f\in\cH$ be an element we want to reconstruct from measurements $c_k=\inner{s_k}{f}$, where $\{s_k\}$ is the set of sampling vectors.
The reconstruction is an approximation of $f$ of the form $\tilde f=\sum_j\beta_jr_j$, where $r_j$ are called reconstruction vectors.
In practice, the index sets for $k$ and $j$ should be finite, which means that the quality of the reconstruction $\tilde f\in \spn_j\{r_j\}$ depends heavily on the choice of reconstruction vectors.
In the following, we assume that the unknown function $f$ is an element of $L^2(\R^d)$ supported in the compact set $E$, and we focus on frequency measurements (i.e.\ the $c_k$’s are certain values of the Fourier transform of $f$)
.

Let $Y \subset \R^d$ be a closed, simply connected set with $0$ in its interior, referred to as our \emph{observation window}.
The set of frequencies where we measure $f$ is denoted $\Omega \subset Y$ and is referred to as a \emph{sampling scheme}.
The \emph{bandwidth} of the observation window/sampling scheme is $\max_{y\in Y} |y|_\infty$ .

If $\Omega$ is not a subset of a (sufficiently nice subset of a) lattice in $\R^d$, the classical theory is not applicable, and we need to use the so-called \emph{non-uniform generalized sampling}, see \cite{Adcock:2014,Adcock:2017}.
In the following, we recall some results from these papers.

\begin{theorem}[\cite{Adcock:2014}]\label{thm:AGH14}
    Assume that we have a countable sampling scheme $\Omega\subset\R^d$, a finite dimensional reconstruction space, i.e., a subspace $T\subset\cH=L^2(\R^d)$, and a non-negative, bounded, selfadjoint operator $S\colon\cH\to\cH$ such that for all $f\in\cH$, $Sf$ is uniquely determined by $\{\hat f(\xi)\}_{\xi\in\Omega}$ and there exists a $C>0$ such that for all $f\in T$, $\inner{Sf}{f}\ge C\norm{f}^2$.
    Then for every $f\in\cH$ there exists a unique $\tilde f=F(f)\in T$ determined by $\forall g\in T\colon \inner{Sf}{g}=\inner{S\tilde f}{g}$ satisfying that $\forall f, h\in\cH\colon \norm{f-F(f+h)}\le\sqrt{\frac{\norm{S}}{C}}(\norm{f-Pf}+\norm{h})$, where $P$ denotes the orthogonal projection on $T$.
\end{theorem}

To apply the theorem, we need to verify the assumptions.
In \cite{Adcock:2017}, it is shown that if $T\subset L^2(E) \subset\cH$ for a sufficiently nice, compactly supported set $E$, and if $\{x\mapsto \sqrt{\mu_\xi} e^{2\pi\ci\xi\cdot x\bbbone_E(x)}\}_{\xi\in\Omega}$ with suitably chosen $\mu_\xi>0$ is a so-called \emph{weighted Fourier frame} for $L^2(E)$, then $S$ can be chosen to be the corresponding \emph{frame operator}, and they establish sufficient conditions on $\Omega$ which ensures that such a weighted frame exists.
Since we want to go to finite subsets of $\Omega$, choose a labelling such that $\{\xi_n\}_{n=1}^\infty = \Omega$ and let $\Omega_N = \{\xi_n\}_{n=1}^N$ for positive integers $N$.
With $T$ and $S$ as above, \cite{Adcock:2017} establishes sufficient conditions for the corresponding \emph{truncated frame operator} $S_N$ to satisfy the conditions of \cref{thm:AGH14}.
Before we can state the condition, we need to introduce some additional notation.
\begin{definition}[$\abs{\cdot}_E$, $E^\circ$]
   Let $E$ be compact, convex and symmetric. Then we define the norm $\abs{\cdot}_E$ by
   $%\begin{equation*}
      \abs{x}_E=\inf\{a>0:x\in aE\}
   $ %\end{equation*}
   and 
   $%\begin{equation*}
      E^\circ = \{y\in\R^d : x\cdot y\le 1, \forall x\in E\}.
   $%\end{equation*}
\end{definition}
Note that $E^\circ$ is automatically also compact, convex and symmetric, so $\abs{\cdot}_{E^\circ}$ is well-defined.
We can now define the relative, \emph{inverse density} of $\Omega$ in an observation window $Y$ with respect to $\abs{\cdot}_{E^\circ}$: $\delta_{E^\circ}(\Omega, Y)=\sup_{y\in Y}\inf_{\xi\in\Omega}\abs{\xi-y}_{E^\circ}$.
If the norm used is just the usual one, we omit the subscript ``$E^\circ$''.
\begin{theorem}[\cite{Adcock:2017}]
    \label{thm:AGH15a}
    If $\delta_{E^\circ}(\Omega, \R^d)<\frac14$, then there exists weights $\mu_{\xi_n}>0$, $\xi_n\in\Omega$ such that $\{e_n\}_{n\in\N}$ is a frame for $L^2(E)$.
    Moreover, the weights $\mu_{\xi_n}$ may be chosen as the measures of the Voronoi regions (of $\xi_n$) with respect to the $\abs{\cdot}_{E^\circ}$ norm.
\end{theorem}
This result is obviously not applicable for finite sets $\Omega$, as the inverse density of a finite set is infinite.
But since we are only interested in vectors in the finite dimensional $T$, we can do with less than a frame for $L^2(E)$:
\begin{theorem}[\cite{Adcock:2017}]
    \label{thm:AGH2015_density}
    Let $T\subset L^2(E)$ be finite-dimensional, $E$ be a compact, convex and symmetric set, and assume $\delta_{E^\circ}(\Omega, \R^d)<\frac14$. 
    Let $\{e_n\}_{n\in\N}$ be a frame satisfying the conditions in \cref{thm:AGH15a} and let $A$ and $B$ be the frame bounds. 
    Assume that $Y$ is an observation window, such that $\Omega_N = \Omega\cap Y$ is finite and that
    \begin{equation*}
        R(\Omega_N, T)
        = \sup\{\smashoperator[r]{\sum_{\xi\in\Omega\setminus\Omega_N}}\mu_\xi\abs{\hat f(\xi)}^2:f\in T, \norm{f}=1\}<A.
    \end{equation*}
    Then $(T, \Omega_N, S_N)$ satisfies the conditions of \cref{thm:AGH14} with $C = A - R(\Omega_N, T)$ and $F(f) = \tilde f = \argmin_{g\in T}\sum_{n=1}^N \mu_{\xi_n} \abs{\hat f(\xi_n) - \hat g(\xi_n)}^2.$
\end{theorem}

The reconstruction is computed by solving a least squares problem where the design matrix is a finite section of the change of basis matrix between the sampling and reconstruction vectors.
The quality and convergence speed of an iterative solver like conjugate gradients depends on the condition number of this matrix \cite{Golub:van_Loan:2013}.
It is the weights in \cref{thm:AGH15a,thm:AGH2015_density} that ensure small condition numbers for non-regular sampling schemes.
In particular, without weights the numerical stability would worsen for sampling schemes with clusters.
A sampling scheme \emph{can} cause large conditions numbers if the density requirement is violated or if the bandwidth is small \cite[Theorem 6.1]{Adcock:2014}.

\section{Binomial, Poisson and determinantal point processes}
\label{sec:point_process}

In this section we assume $\Omega$ is a locally finite spatial point process in $\R^d$ without multiple points and any accumulation point, i.e.\ we can view $\Omega$ as a closed random subset of $\R^d$ so that with probability one, for any bounded set $B\subset\R^d$, the intersection $\Omega\cap B$ is finite. For measure theoretical details, see \cite{Moeller:Waagepetersen:2004} and the references therein. We focus on three models: 
binomial, Poisson and determinantal point processes. We 
discuss only the definitions and properties of these models which become relevant for our purpose. 
In particular, in the case where $\Omega$ is a stationary Poisson process, we establish a lower bound on the probability that $\delta(\Omega,Y)<1/4$.

We use the generic notation $K$ for a Borel set $K\subset\R^d$ of finite Lebesgue measure $|K|$. Moreover, $N(K)$ denotes the random number of points in $\Omega\cap K$.

\subsection{Binomial and Poisson processes}

For a given Borel set $K\subset\R^d$ with $0<|K|<\infty$, if $\Omega\cap K$ consists of a fixed number $N(K)=n>0$ of independent points $\xi_1,\ldots,\xi_n$ which are uniformly distributed on $K$ (i.e.\ $\mathrm P(\xi\in A)=|A|/|K|$ for Borel sets $A\subseteq K$), then $\Omega\cap K=\{\xi_1,\ldots,\xi_n\}$ is called a binomial point process. 
This is the model used in \cite{Bass:Groechenig:2005}, and it can be extended as follows. Let $\rho>0$. If for any Borel set $K\subset\R^d$ with $|K|<\infty$, we have that 
$N(K)$ is Poisson distributed with parameter $\lambda=\rho|K|$, and conditional on $N(K)=n>0$, the $n$ points in $\Omega\cap K$ form a binomial point process, 
then $\Omega$ is a stationary Poisson process on $\R^d$ with intensity $\rho$. The process is often referred to as ``complete spatial randomness''. % because of the lack of interaction or dependence properties. 

The following theorem is verified in \ref{sec:poisson_prob_proof}, and it is of particular interest for us when $K=Y$ is the observation window.
\begin{theorem}
    \label{t:5}
    Suppose $\Omega$ is a stationary Poisson process with intensity $\rho>0$ and consider any Borel set $K\subset\R^d$ with $|K|<\infty$. Then
    \begin{align}\label{e:Presult}
        &\mathrm P(\delta(\Omega,K)<1/4)\ge
        \nonumber
        \\
        & 1 - \rho^{d+1}|K|
        \frac{\Gamma\left(\frac{d^2+1}{2}\right)}{\Gamma\left(\frac{d^2}{2}\right)}
        \left\{\frac{\Gamma\left(\frac{d}{2}\right)}{\Gamma\left(\frac{d+1}{2}\right)}\right\}^d
        \frac{\Gamma\left(\frac{2}{2}\right)\cdots\Gamma\left(\frac{d}{2}\right)}
        {\Gamma\left(\frac{1}{2}\right)\cdots\Gamma\left(\frac{d-1}{2}\right)}
        \frac{1}{d\left(\rho\omega_d\right)^d}\Gamma\left(d,\rho\omega_d4^{-d}\right).
    \end{align} 
\end{theorem}

\subsection{Determinantal point processes}

Determinantal point processes (DPPs) are models for repulsiveness (inhibition or regularity) between points in ``space'', where space means $\R^d$ in the present setting. DPPs are of interest because of their applications in mathematical physics, combinatorics, random-matrix theory, machine learning and spatial statistics, and because they provide rather flexible and tractable parametric models, see \cite{Lavancier:Moeller:Rubak:2014} and the references therein. 

Recall that $\rho^{(n)}: \R^{dn}\mapsto[0,\infty)$, $n=1,2,\ldots$, are so-called 
joint intensities for $\Omega$ if
 for any pairwise disjoint bounded Borel sets $A_1,\ldots,A_n\subseteq\R^d$, 
\[\mathrm E\{N(A_1)\cdots N(A_n)\}=\int_{A_1}\cdots\int_{A_n}\rho^{(n)}(x_1,\ldots,x_n)\,\mathrm dx_1\ldots\,\mathrm dx_n\]
and this mean value is finite.
 Note that $\rho^{(n)}(x_1,\ldots,x_n)$ is only uniquely determined for Lebesgue almost all $(x_1,\ldots,x_n)\in\R^d\times\ldots\times\R^d$ ($n$ times). For infinitesimally small $A_1,\ldots,A_n$ containing the points $x_1,\ldots,x_n$, intuitively $\rho^{(n)}(x_1,\ldots,x_n)|A_1|\cdots|A_n|$ is the probability that $\Omega\cap|A_i|\not=\emptyset$, $i=1,\ldots,n$.
Now, $\Omega$ is said to be a DPP with kernel $C:\R^d\times \R^d\mapsto\mathbb C$ if we can take $\rho^{(n)}(x_1,\ldots,x_n)
    = {\mathrm{det}}\{C(x_i,x_j)\}_{i,j=1,\ldots,n}$ 
for all $n=1,2,\ldots$ and $(x_1,\ldots,x_n)\in\R^d\times\ldots\times\R^d$.

The stationary Poisson process with fixed intensity $\rho$ is the very special case where $\rho^{(n)}(x_1,\ldots,x_n)=\rho^n$, i.e.\ when $C(x,x)=\rho$ and $C(x,y)=0$ whenever $x\not=y$. 

In this paper we restrict attention to kernels defined by a continuous complex function $C_0\in L^2(\R^d)$ so that $C(x,y)=C_0(x-y)$ is assumed to be positive semi-definite. % (then $C$ is called a stationary covariance function). 
As discussed in \cite{Lavancier:Moeller:Rubak:2014} these are rather mild conditions, though the continuity assumption excludes the case of the Poisson process. For $\varphi\in L^1(\R^d)$, denote its inverse Fourier transform by $\mathcal F^{-1}(\varphi)$, i.e.\
\[\mathcal F^{-1}(\varphi)(x)=\int \varphi(y)\exp(2\pi ix\cdot y)\,\mathrm dy,\qquad x\in\R^d. \] 
Then the existence of the DPP is  equivalent to the existence of a function $\varphi\in L^1(\R^d)$ such that $0\le\varphi\le1$ and $C_0=\mathcal F^{-1}(\varphi)$, cf.\ Corollary~3.3 in \cite{Lavancier:Moeller:Rubak:2014}. Then $\varphi$ is called the spectral density of the DPP, and the intensity $\rho^{(1)}(x)=\rho$ does not depend on $x$ and is given by $\rho=C_0(0)$. 
Moreover, by comparison with the Poisson process, the repulsiveness of the process is reflected by that fact that 
$\rho^{(n)}(x_1,\ldots,x_n)\le\rho^n$ (see \cite{Lavancier:Moeller:Rubak:2014}). Therefore, 
we expect that $\mathrm P(\delta(\Omega,K)<1/4)$ is larger than in the
Poisson case. In fact this in accordance with our experimental results discussed in \cref{e:experiments}.
Moreover, an important observation in \cite{Lavancier:Moeller:Rubak:2014} is the trade-off between how large $\rho$ can be and how strong the repulsiveness in the DPP can be. 

Several examples of such kernels $C$ %stationary covariance functions (Gaussian, Whittle-Mat{\'e}rn, generalized Cauchy and power exponential spectral model) 
and their spectral densities are discussed in \cite{Lavancier:Moeller:Rubak:2014}. A simulation algorithm for the corresponding DPPs is available in the \texttt{spatstat} package \cite{Baddeley:Rubak:Turner:2015} --- in particular with the contribution from \cite{Lavancier:Moeller:Rubak:2014}.

For instance, a power exponential spectral model is specified by its spectral density given by
\[\varphi(x)=\rho\left(\alpha/\alpha_{\max}\right)^d
\exp\left(-\|\alpha x\|^\nu\right),\qquad x\in\R^d,\]
where $\|\cdot\|$ denotes usual Euclidean distance, $\rho>0$ is the intensity, $\nu>0$ is a shape parameter and $\alpha\in(0,\alpha_{\max}]$ is a scale parameter, where 
\[\alpha_{\max}=\sqrt{\pi}\left\{\frac{\Gamma(d/\nu+1)}{\Gamma(d/2+1)\rho}\right\}^{1/d}.\]
For fixed values of $\rho$ and $\nu$, the DPP becomes more and more repulsive as $\alpha$ increases to $\alpha_{\max}$; and for a fixed value of $\rho$ and letting $\alpha=\alpha_{\max}$, as $\nu$ increases, the DPP ranges from the Poisson process (the limiting case of $\nu\rightarrow0$) to 
the ``most repulsive DPP'' (the limiting case of $\nu\rightarrow\infty$), see \cite{Lavancier:Moeller:Rubak:2014} for the details.

\section{Experiments}\label{e:experiments}

We compare the condition numbers obtained when using random sampling schemes generated from the binomial, Poisson and determinantal point processes.
The reconstruction functions are Daubechies scaling functions (see e.g.\ \cite{Hernandez:Weiss:1996}) on $[-\tfrac12, \tfrac12]^2$ using the boundary correction of \cite{Cohen:Daubechies:Vial:1993}. The scaling functions are chosen for convenience due to the fact that the software from \cite{Jacobsen:Nielsen:Rasmussen:2016} reconstructs scaling functions instead of wavelets. However, this does not affect the numerical stability of generalized sampling as this only depends on the reconstruction subspace and not on the basis of choice, see also \cite[§4.4]{MR3175085}.

\begin{figure}
	\centering
	\includegraphics[height=7cm]{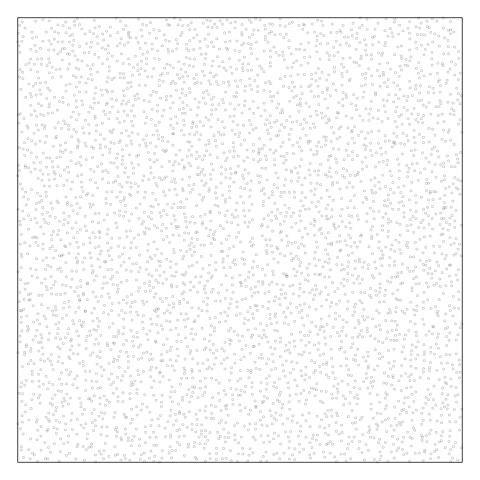}
	\includegraphics[height=7cm]{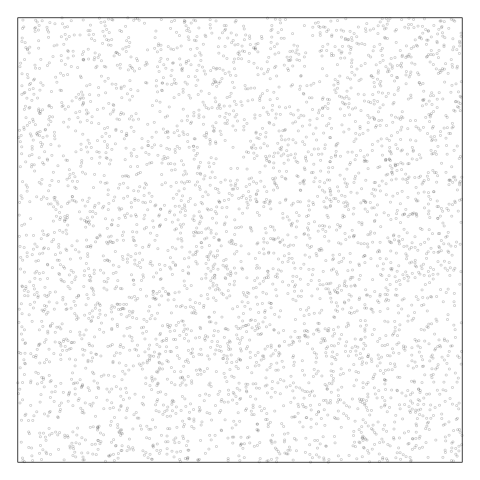}
    \caption{A realization of a DPP (left) and a Poisson process (right) on the observation window $[-64, 64]^2$.}
	\label{fig:PP_simulation_example}
\end{figure}

\cref{fig:PP_simulation_example} shows a simulation from a DPP and a Poisson process when we have the same mean number of points in an observation window.
The Poisson (and binomial) point processes show ``complete spatial randomness'', whereas the DPP is repulsive and give rise to a more regular point pattern with less ``empty space''.

Let the observation window be $Y = [-64,64]^2$, since the theory of generalized sampling dictates that the sampling area should be symmetric with a bandwidth that depends on the number of functions we wish to reconstruct.
We let the DPP kernel be a power exponential spectral model with shape parameter $\nu=10$ and scale parameter $\alpha=\alpha_{\max}$; this is very close to the ``most repulsive DPP'' (see \cite{Lavancier:Moeller:Rubak:2014}).
 % Thus the Poisson process and the DPP are specified by their intensity parameter $\rho$.

For each point process model we made 100 simulations, where the fixed number of points in the binomial point process is $n=4096$ and the intensity of both the Poisson process and the DPP is $\rho=1/4$.  Thus in all cases, the expected number of points is $\mathrm E\{N(Y)\}=4096$. 
Simulations of the point processes were performed in R \cite{R} with the \texttt{spatstat} package \cite{Baddeley:Rubak:Turner:2015}.
The condition numbers were computed in Julia \cite{Bezanson:Edelman:Karpinski:Shah:2014} with the GeneralizedSampling package from \cite{Jacobsen:Nielsen:Rasmussen:2016}.
The code used in the experiments is available at \url{http://doi.org/10.5281/zenodo.887928}.

Depending on the nature of the sampling locations and the choice of reconstruction scaling function, we can recover \emph{up to} $32^2 = 1024$ scaling functions, i.e., scaling functions at scale 5. 
However, this is overly optimistic in our experiments with non-uniform points that do not fulfill the density assumption of \cref{thm:AGH2015_density}.
The support of the Daubechies $p$ scaling functions at scale $J$ is of length $2^{-J}(2p-1)$.
Since our reconstruction space is $L^2([-\tfrac12, \tfrac12])$ we must have less than or equal to 7 vanishing moments of the reconstructed wavelet to ensure that the support of the scaling functions are fully contained in $[-\tfrac12, \tfrac12]$.

\subsection{Results}

We investigated three questions: 
\begin{enumerate*}[label=(\arabic*)]
    \item Which Daubechies basis is best for reconstruction (independently of the choice of point process)?
    \item Which point process is best for reconstruction (independently of the choice of Daubechies basis)?  
    \item How does the density of a point pattern influence the reconstruction?
\end{enumerate*}

For all three point processes, the Haar scaling functions gave rise to the lowest condition numbers. In fact, as shown in Table \ref{tab:cond} the condition numbers for Haar was measured in tens, where the condition numbers for the other functions were measured in hundreds or thousands. This is consistent with the theoretical results for generalized sampling with regular sampling schemes in \cite{Adcock:Hansen:Kutyniok:Ma:2014}.

\begin{table}[h]
\begin{center}
	\begin{tabular}{ *{4}{c} }
		\toprule
      		 & Binomial         & Poisson          & Determinantal
		\\
		\cmidrule{2-4}
		Haar & 51.4 (1.4)       & 54.4 (1.6)       & 40.2 (0.9)
		\\
		db2  & 371.3 (15.4)     & 382.5 (17.3)     & 278.3 (9.3)
		\\
		db3  & 1272.2 (69.3)    & 1319.2 (76.9)    & 905.5 (33.9)
		\\
		db4  & 4658.0 (313.8)   & 4378.1 (261.8)   & 3026.0 (129.8)
		\\
		db5  & 10210.1 (672.6)  & 9962.0 (649.8)   & 6736.8 (287.3)
		\\
		db6  & 22545.3 (1845.6) & 20396.0 (1425.6) & 13183.9 (650.6)
		\\
		db7  & 41934.9 (4152.0) & 39004.6 (3028.6) & 23390.8 (1204.0)
		\\
		\bottomrule
	\end{tabular}
\end{center}
       \caption{Average condition numbers and their estimated standard deviations (in parentheses) obtained for the first 6 Daubechies scaling functions.}
\label{tab:cond}
 \end{table}

As can be concluded from Table~\ref{tab:cond}, the DPP consistently gave rise to the lowest condition numbers and standard deviations. 
Strong evidence suggested that the mean condition number for the DPP is significantly lower than for both the binomial and Poisson process sampling schemes (for the Haar scaling functions that generally have the lowest condition numbers, the p-value is $<10^{-6}$ when using Tykey's range test).
For both the binomial and Poisson process sampling schemes the 
average condition numbers and standard deviations were similar.

\begin{figure}[h]
\centering
\includegraphics{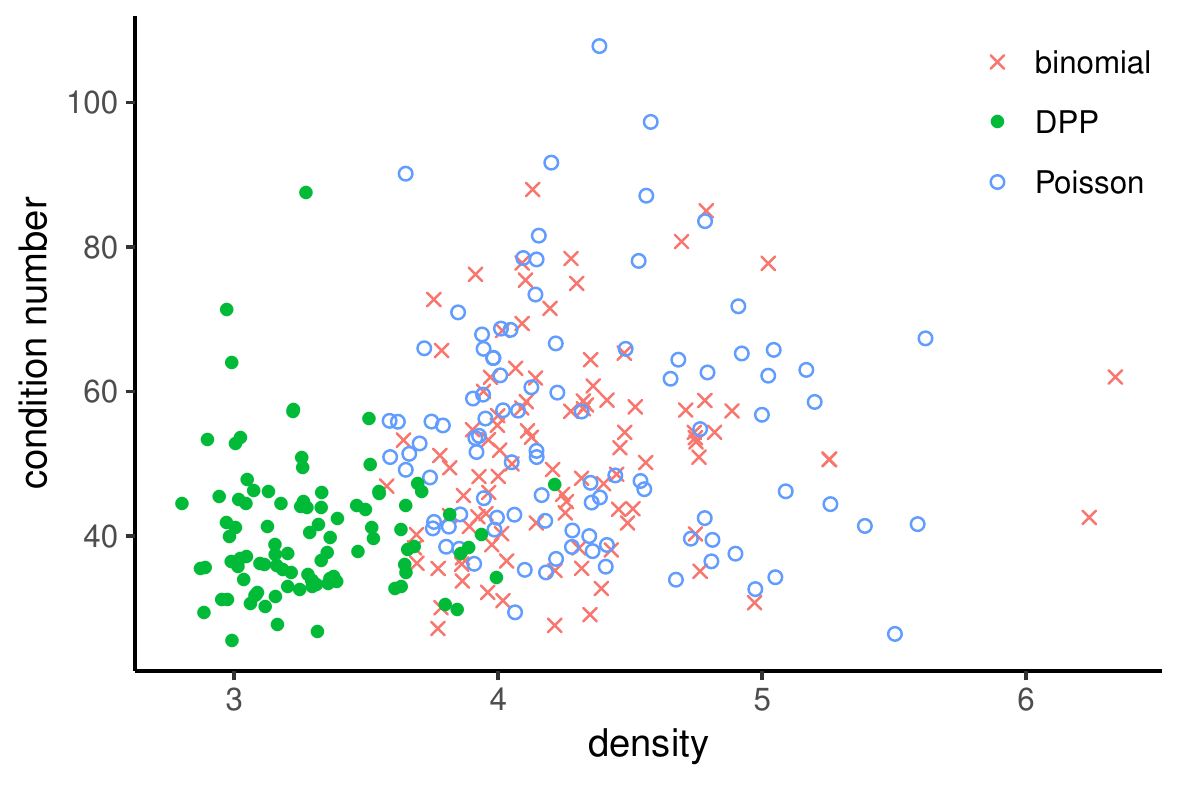}
\caption{The density of  sampling schemes versus the condition number obtained using the Haar scaling function.}
\label{fig:density_cond}
\end{figure}

As mentioned in \cref{thm:AGH2015_density}, a small density of a point pattern is (part of) a sufficient condition for a small condition number. 
In our simulations \emph{none} of the realizations had small densities.
\cref{fig:density_cond} shows scatterplots of densities of the point patterns versus the condition number for reconstruction with the Haar scaling functions, where for each of the processes \emph{individually}, there is no significant correlation between the density and the condition numbers ($\pval \approx 0.17$ for the binomial point process, $\pval \approx 0.50$ for the Poisson process and $\pval \approx 0.72$ for the DPP  using the test based on  Pearson's product-moment correlation). 

This is in contrast to the case of deterministic sampling, where a violation of the density condition has severe impact on the condition number. An example is given in Table~\ref{tab:determin} showing the rapidly increasing condition numbers obtained for the Haar function and deterministic sampling on $\varepsilon\mathbb{Z}^2$ for increasing values of $\varepsilon$ using the same observation window as above and recovery of scaling functions up to level 5. 
\begin{table}[h]
\begin{center}
	\begin{tabular}{l *{5}{c} }
		\toprule
		$\varepsilon$    & 1    & 1.25  & 1.5   & 1.75   & 2
		\\
		\midrule
		Condition number & 1.11 & 59.59 & 83.64 & 274.49 & $1.6 \cdot 10^{20}$
		\\ 
		\bottomrule
      \end{tabular}
\end{center}
\caption{Condition numbers for regular sampling on a subset of $\varepsilon\mathbb{Z}^2$.}
\label{tab:determin}
\end{table}

\section{Conclusion}

In this paper we have investigated how the choice of sampling according to a random sampling scheme affects the numerical stability of 2D non-uniform generalized sampling measured by the condition number.

We have compared three kinds of random point patterns, generated by binomial, Poisson or determinantal point processes.
As reconstruction bases we have considered Daubechies scaling functions with polynomial reconstruction of degree 1 through 7 and the moment preserving boundary correction of \cite{Cohen:Daubechies:Vial:1993}.

Our results can be summarized as follows: 
\begin{enumerate*}[label=\arabic*)]
    \item We have shown that certain sampling schemes can give rise to large densities while retaining small  condition numbers. 
        This is surprising because the theory dictates that a small density is part of a sufficient condition for small a condition number.
	\item For all scaling functions sampling points from the determinantal point process yields the lowest mean condition number.
	\item\label{item:conclusion_wavelet} For all kinds of sampling points the Haar scaling function yields the lowest condition numbers.
\end{enumerate*}
Point \cref{item:conclusion_wavelet} is consistent with the findings of \cite{Adcock:Hansen:Kutyniok:Ma:2014}.

A theoretical result has been obtained for Poisson processes in \cref{t:5} and it would be interesting to extend the validity to the case of a general DPP. However, the  the proof of Theorem~\cref{t:5} relies heavily on the strong independence properties of a Poisson process, and a much more sophisticated approach will be needed to extend the result to the case of a general DPP. We leave this  for future research.

\appendix

\section{Proof of \cref{t:5}}
\label{sec:poisson_prob_proof}

Let $\Omega$ be a stationary Poisson process on $\R^d$ with intensity $\rho>0$. 
Many statements in the sequel %this appendix 
are only satisfied almost surely (a.s.), i.e.\ with probability one. 

We refer to the points of $\Omega$ as nuclei and consider the Voronoi tessellation with cells generated by the nuclei, i.e.\ the Voronoi cell associated to a nucleus consists of all points in $\R^d$ which are at least as close to that cucleus as to the other nuclei (with respect to usual distance in $\R^d$). 
With probability one each vertex in the Voronoi tessellation is given by intersection of $d+1$ Voronoi cells. 
If $\{\xi_0,\ldots,\xi_d\}$ is the set of the corresponding nuclei, they define a.s.\ a unique $d$-dimensional closed ball $B(\xi_0,\ldots,\xi_d)$ containing $\{\xi_0,\ldots,\xi_d\}$ in its boundary. 
Its center $C(\xi_0,\ldots,\xi_d)$ is then the Voronoi vertex. 
Conversely, for each set $\{\xi_0,\ldots,\xi_d\}$ of $d+1$ pairwise distinct nuclei such that $B(\xi_0,\ldots,\xi_d)\cap\Omega=\{\xi_0,\ldots,\xi_d\}$, $C(\xi_0,\ldots,\xi_d)$ is a.s.\ a Voronoi vertex. 
We denote $R(\xi_0,\ldots,\xi_d)$ the radius of $B(\xi_0,\ldots,\xi_d)$. 
Let $K\subset\R^d$ be a Borel set of finite Lebesgue measure $|K|$. 
Considering only those Voronoi vertices which are contained in $K$, the largest nuclei-vertex distance is a.s.\ given by
\begin{align*}
	R(\Omega,K) = \max\{R(\xi_0,\ldots,\xi_d):&\,
		\{\xi_0,\ldots,\xi_d\}\subset\Omega\mbox{ is of cardinality $d+1$},
		\\
		&B(\xi_0,\ldots,\xi_d)\cap\Omega=\{\xi_0,\ldots,\xi_d\},\ 
	C(\xi_0,\ldots,\xi_d)\in K\}
\end{align*}
noticing that the number of such vertices is a.s.\ finite.
We want to estimate the probability $p = \mathrm P(R(\Omega,K)<1/4)$. 

We have
\begin{align*}
	1-p = \mathrm P\bigg(&\exists \{\xi_0,\ldots,\xi_d\}\subset\Omega\mbox{ of cardinality
	$d+1$}:\ B(\xi_0,\ldots,\xi_d)\cap\Omega=\{\xi_0,\ldots,\xi_d\},\\
	& C(\xi_0,\ldots,\xi_d)\in K,\ R(\xi_0,\ldots,\xi_d)\ge 1/4\bigg)
\end{align*}
and this is at most the mean value
\begin{align*}
	I=\mathrm E
	&\sum_{\{\xi_0,\ldots,\xi_d\}\subset\Omega\mbox{ of cardinality
	$d+1$}}\mathbf1\bigg[
		B(\xi_0,\ldots,\xi_d)\cap\Omega=\{\xi_0,\ldots,\xi_d\},\\
	&C(\xi_0,\ldots,\xi_d)\in K,\ R(\xi_0,\ldots,\xi_d)\ge 1/4\bigg]
\end{align*}
where $\mathbf1\left[\cdot\right]$ denotes the indicator function. 
Since $\Omega$ is a stationary Poisson
point process we can evaluate $I$: Let
	$\omega_d=\frac{\pi^{d/2}}{\Gamma\left(1+\frac{d}{2}\right)}$ 
and $ 
	\sigma_d=\frac{2\pi^{d/2}}{\Gamma\left(\frac{d}{2}\right)}$
be the volume respective surface area 
of the $d$-dimensional unit ball.
By the extended Slivnyak-Mecke theorem (see \cite{Moeller:Waagepetersen:2004}),
\begin{align*}
	I= \rho^{d+1}\int\cdots\int
	&\,\mathbf 1 \left[
	C(\xi_0,\ldots,\xi_d)\in K,\ R(\xi_0,\ldots,\xi_d)\ge 1/4\right]\\
	&\,\mathrm P\left(B(\xi_0,\ldots,\xi_d)\cap\Omega=\emptyset\right)
	\,\mathrm d\xi_0\cdots\,\mathrm d\xi_d
\end{align*}
where $\mathrm P\left(B(\xi_0,\ldots,\xi_d)\cap\Omega=\emptyset\right)
	= \exp\left(-\rho R(\xi_0,\ldots,\xi_d)^d\right)$.

To evaluate this integral we shift coordinates from
$(\xi_0,\ldots,\xi_d)$ to $(c,r,u_0,\ldots,u_d)$, where
$c=C(\xi_0,\ldots,\xi_d)$, $r=R(\xi_0,\ldots,\xi_d)$, and
$x_i=c+ru_i$, $i=0,\ldots,d$. Let $\nabla=\nabla(u_0,\ldots,u_d)$ be
$d!$ times the Lebesgue measure of the simplex with vertices
$u_0,\ldots,u_d$. Then 
by the Blaschke-Petkantschin's formula (see \cite{Moeller:1989} and the references therein),
	$\mathrm d\xi_0\cdots\,\mathrm d\xi_d
	= \sigma_d^{d+1}\nabla r^{d^2-1} \,\mathrm dc\,\mathrm dr\,\mathrm du_0\cdots\,\mathrm du_d$,
where $\mathrm dc$ is Lebesgue measure on $\R^d$,
$\mathrm dr$  is Lebesgue measure on $(0,\infty)$, and 
$\mathrm du_i$ is the uniform distribution on the unit sphere in $\R^d$.
Thus
\begin{align*}
	I=&\,\rho^{d+1}|K|\int_{1/4}^\infty r^{d^2-1}
	\exp\left(-\rho r^d\right)\,\mathrm dr\,\int\cdots\int\nabla\,\mathrm du_0\cdots\,\mathrm du_d\,.
\end{align*}
\noindent Here
\begin{equation*}
	\int_{1/4}^\infty r^{d^2-1}
	\exp\left(-\rho\omega_d r^d\right)\,\mathrm dr=
	\frac{1}{d\left(\rho\omega_d\right)^d}\int_{\rho\omega_d4^{-d}}^\infty
	t^{d-1}\mathrm e^{-t}\,\mathrm dt=
	\frac{1}{d\left(\rho\omega_d\right)^d}\Gamma\left(d,\rho\omega_d4^{-d}\right)
\end{equation*}
where by integration by parts $\Gamma(d,s)=(d-1)\Gamma(d-1,s)+s^{d-1}\mathrm e^{-s}$, $s>0$.
Furthermore,
\begin{equation*}
	\int\cdots\int\nabla\,\mathrm du_0\cdots\,\mathrm du_d=
	\frac{\Gamma\left(\frac{d^2+1}{2}\right)}{\Gamma\left(\frac{d^2}{2}\right)}
	\left\{\frac{\Gamma\left(\frac{d}{2}\right)}{\Gamma\left(\frac{d+1}{2}\right)}\right\}^d
	\frac{\Gamma\left(\frac{2}{2}\right)\cdots\Gamma\left(\frac{d}{2}\right)}
	{\Gamma\left(\frac{1}{2}\right)\cdots\Gamma\left(\frac{d-1}{2}\right)}
\end{equation*}
which reduces to 1 if $d=1$ (see \cite{Moeller:1989} and the references therein).
Consequently,
\begin{align*}
	&\mathrm P(R(\Omega,K)<1/4)\ge\\
	&1-\rho^{d+1}|K|
	\frac{\Gamma\left(\frac{d^2+1}{2}\right)}{\Gamma\left(\frac{d^2}{2}\right)}
	\left\{\frac{\Gamma\left(\frac{d}{2}\right)}{\Gamma\left(\frac{d+1}{2}\right)}\right\}^d
	\frac{\Gamma\left(\frac{2}{2}\right)\cdots\Gamma\left(\frac{d}{2}\right)}
	{\Gamma\left(\frac{1}{2}\right)\cdots\Gamma\left(\frac{d-1}{2}\right)}
	\frac{1}{d\left(\rho\omega_d\right)^d}\Gamma\left(d,\rho\omega_d4^{-d}\right)\,.
\end{align*}

\subsection*{Acknowledgments}

Supported by the Danish Council for Independent Research | Natural Sciences, grant 12-124675, "Mathematical and Statistical Analysis of Spatial Data".
Supported by the Centre for Stochastic Geometry and Advanced Bioimaging, funded by a grant (8721) from the Villum Foundation.

The authors would like to thank Ege Rubak for practical help with the \texttt{spatstat} package.

\printbibliography

\end{document}